\documentclass[]{spie}  

\usepackage[]{graphicx}

\usepackage{amsmath,amssymb,amsfonts,mathrsfs} 
\usepackage{subfigure}
\usepackage{fancyhdr} 
\usepackage{lastpage} 
\usepackage{url} 
\usepackage{array}
\usepackage{tabularx}
\usepackage{supertabular}
\usepackage{multirow}
\usepackage{rotating}
\usepackage{nomencl}
\usepackage{bm}
\usepackage{amsmath}
\usepackage{calrsfs}
\usepackage{bbm}
\usepackage{epstopdf}

\usepackage{color} 
\usepackage{epsfig}

\title{Focal-plane wavefront sensing with high-order adaptive optics systems}

\author{Visa Korkiakoski\supit{a,b}, Christoph U. Keller\supit{b}, Niek Doelman\supit{c,b}, Matthew Kenworthy\supit{b}, Gilles Otten\supit{b} and Michel Verhaegen\supit{a}
\skiplinehalf
\supit{a}Delft Center for Systems and Control, Mekelweg 2, 2628CD Delft, The Netherlands \\
\supit{b}Leiden Observatory, Leiden University, P.O. Box 9513, 2300 RA Leiden, The Netherlands \\
\supit{c}TNO Science and Industry, Stieltjesweg 1, 2628CK Delft, The Netherlands \\
}

\authorinfo{Further author information: (Send correspondence to V.K.)\\
V.K.: E-mail: v.a.korkiakoski@tudelft.nl}

\begin{document} 
\maketitle

\newcommand{\josa}{JOSA}
\newcommand{\apj}{APJ}
\newcommand{\ao}{Applied Optics}
\newcommand{\mnras}{ Mon. Not. R. Astron. Soc.}
\newcommand{\amp}{\&}
\newcommand{\pasp}{Publ. Astron. Soc. Pac.}
\newcommand{\aeta}{Astron. Astrophys.}
\newcommand{\aap}{\aeta} 
\newcommand{\aaps}{Astronomy and Astrophysics Supplement}
\newcommand{\josaa}{Journal of the Optical Society of America A}
\newcommand{\optcom}{Optics Communications}
\newcommand{\aplopt}{Applied Optics}
\newcommand{\optlet}{Optics Letters}
\newcommand{\solphys}{Solar Physics}

\newcommand{\ft}[1]{\mathscr{F}\left\{#1\right\}}
\newcommand{\ift}[1]{\mathscr{F}^{-1}\left\{#1\right\}}
\newcommand{\ftml}[1]{\mathscr{F}\big\{#1\big\}}
\newcommand{\npd}{M}
\newcommand{\nwfs}{M'}
\newcommand{\conv}{\ast}
\newcommand{\half}{0.5}
\newcommand{\quar}{0.25}

\newcommand{\xx}{\alpha}
\newcommand{\yy}{\beta}
\newcommand{\xda}{\alpha_{d1}}
\newcommand{\yda}{\beta_{d1}}
\newcommand{\xdb}{\alpha_{d2}}
\newcommand{\ydb}{\beta_{d2}}
\newcommand{\vrda}{\real{\ft{A_e \phi_{d1}}}}
\newcommand{\vida}{\imag{\ft{A_o \phi_{d1}}}}
\newcommand{\yrda}{\imag{\ft{A_e \phi_{d1}}}}
\newcommand{\yida}{\real{\ft{A_o \phi_{d1}}}}
\newcommand{\vrdb}{\real{\ft{A_e \phi_{d2}}}}
\newcommand{\vidb}{\imag{\ft{A_o \phi_{d2}}}}
\newcommand{\yrdb}{\imag{\ft{A_e \phi_{d2}}}}
\newcommand{\yidb}{\real{\ft{A_o \phi_{d2}}}}
\newcommand{\wfc}{\theta}
\newcommand{\pupipl}{E}
\newcommand{\focapl}{e}
\newcommand{\snd}{\xi}
\newcommand{\reqsupport}{N_\text{arr}}
\newcommand{\pupdim}{N_\text{pup}}
\newcommand{\regpa}{\epsilon}
\newcommand{\vest}{v_s}

\newcommand{\imag}[1]{\text{Im}\left\{#1\right\}}
\newcommand{\real}[1]{\text{Re}\left\{#1\right\}}
\newcommand{\sign}{\text{sign}}

\newcommand{\xph}{\mathbbm x}
\newcommand{\xv}{x}
\renewcommand{\bm}{\mathbbm b}
\newcommand{\bv}{b}
\newcommand{\aft}{A}
\newcommand{\xlim}{\mathbbm X}
\newcommand{\xliv}{X}
\newcommand{\phdm}{\Phi}
\newcommand{\phdv}{\phi}
\newcommand{\ali}{\mathcal A}

\newcommand{\fft}[1]{\mathcal F\left\{#1\right\}}
\newcommand{\ifft}[1]{\mathcal F^{-1}\left\{#1\right\}}


\begin{abstract}
We investigate methods to calibrate the non-common path aberrations at an adaptive optics system having a wavefront-correcting device working at an extremely high resolution (larger than 150$\times$150). We use focal-plane images collected successively, the corresponding phase-diversity information and numerically efficient algorithms to calculate the required wavefront updates. The wavefront correction is applied iteratively until the algorithms converge. Different approaches are studied. In addition of the standard Gerchberg-Saxton algorithm, we test the extension of the Fast \& Furious algorithm that uses three images and creates an estimate of the pupil amplitudes. We also test recently proposed phase-retrieval methods based on convex optimisation. The results indicate that in the framework we consider, the calibration task is easiest with algorithms similar to the Fast \& Furious.

\end{abstract}

\section{Introduction}

Future adaptive optics (AO) systems, for instance in the high-contrast imaging with the extremely large telescopes, will correct the wavefront in increasingly high resolution. The current high-density deformable mirrors (DM) have actuator arrays of up to  $64\times 64$ \cite{poyneer2011} while the largest envisioned AO system at the E-ELT has a DM of $200\times 200$ actuators \cite{verinaud2010,korkiakoski2010}.

To successfully use such high-order AO systems, it is necessary to accurately calibrate the non-common path aberrations caused by the differences in the optical paths between wavefront sensor and the focal-plane scientific camera.

The most popular method for the focal-plane wavefront sensing is perhaps the Gerchberg-Saxton (GS) error reduction algorithm \cite{gerchberg1972, fienup82} and their variations, for instance \cite{green2003, burruss2010}. These are numerically very efficient algorithms, and it is easy to modify them for different applications. However, they suffer from accuracy, in particular because their iterative improvement procedure often stagnates at a local minimum. Various alternatives have been proposed, and a popular approach is to use general numerical optimization techniques  to minimize an error function;  examples include \cite{sauvage2007, riaud2012, paul2013}. However, when the number of optimization parameters is increased, the computational requirements generally rise unacceptably fast.

The numerical issues can be significantly reduced, if the unknown wavefront is sufficiently small. The assumption is valid, for example, when calibrating the non-common path aberrations. This has been exploited in numerous recent works, for instance \cite{gonsalves2001,giveon2007oe, meimon2010, martinache2013, smith2013}. It has also been shown that the Fast \& Furious approach is suitable for effectively controlling wavefront correctors with 20~000--30~000 degrees of freedom \cite{korkiakoski2014}.

In this paper, we investigate different options to do focal-plane wavefront sensing for AO systems with an extremely high-resolution wavefront corrector -- having dimensions larger than $150\times 150$ correction elements. Section \ref{sec:algo} discusses the options: FF-GS, GS and phase-retrieval based on convex optimisation. Sections \ref{sec:results} and \ref{sec:conclusions} show the results and conclusions respectively.

\section{Algorithms}
\label{sec:algo}

The algorithms we consider here can be divided into two groups. First, we consider algorithms that take advantage of the feedback loop when applying the correction: FF and FF-GS. Their performance will be compared to the conventional phase-retrieval setting where the unknown wavefront is retrieved from a set of images -- but with no emphasis of iterative nature of the wavefront correction. The second group includes the traditional Gerchberg-Saxton algorithm and its Fienup variation \cite{gerchberg1972, fienup82}  and a newer class of phase retrieval based on convex optimization.

\subsection{Fast \& Furious}
\label{sec:ff}

\begin{figure*}[hbtp]  \center
\includegraphics[width=\textwidth]{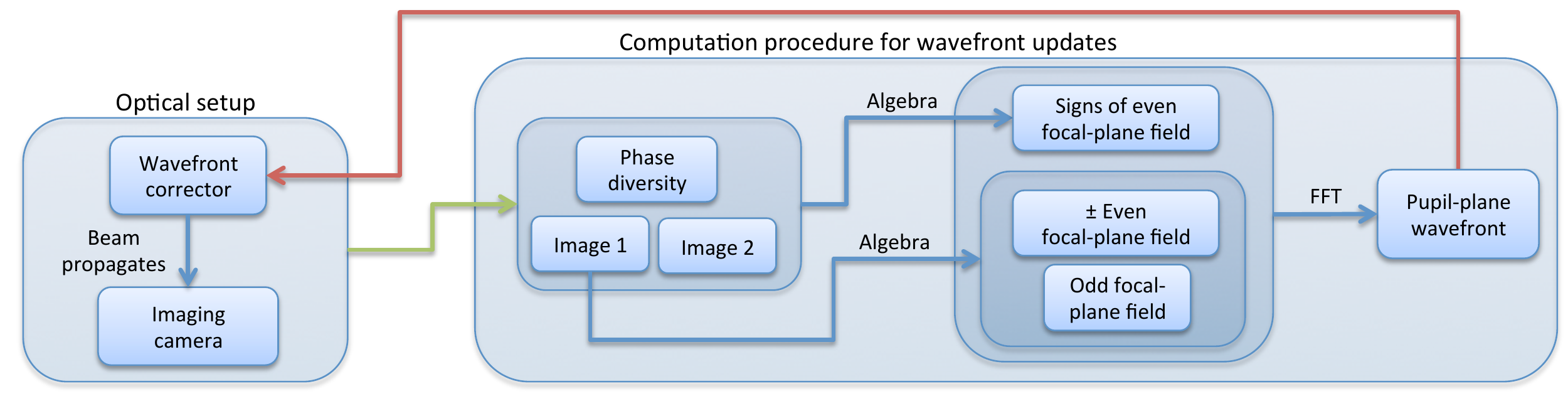} 
\includegraphics[width=\textwidth]{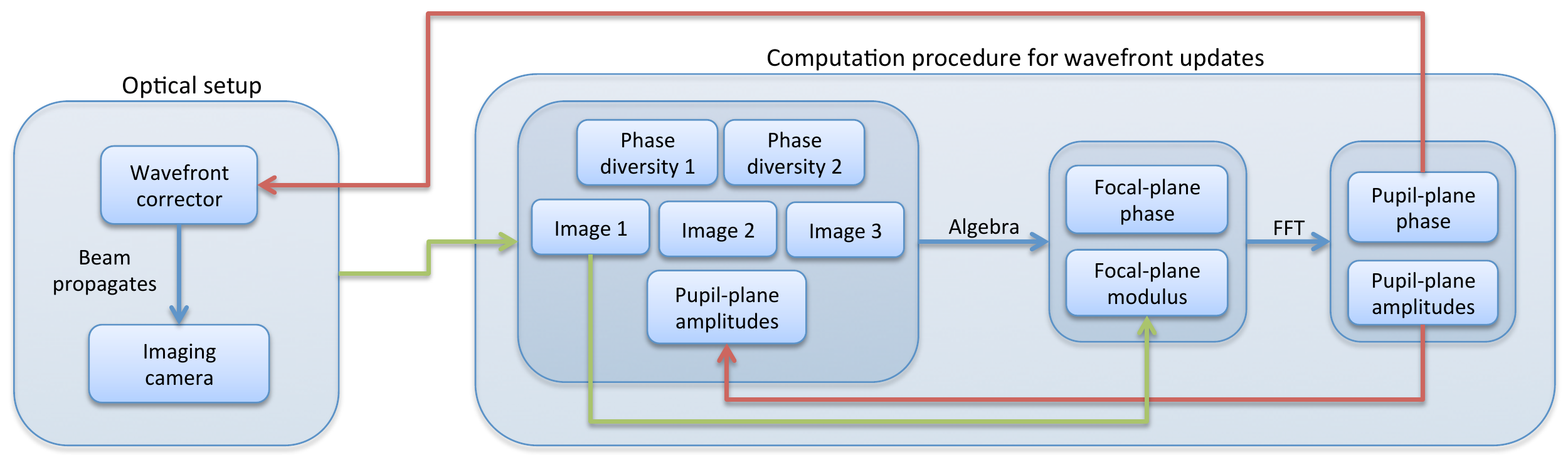} 
\caption{Schematic illustration of the FF and FF-GS algorithms.}
\label{fg:algoscemaff}
\end{figure*}

The Fast \& Furious algorithm is based on iteratively applying a
weak-phase approximation of the wavefront. The main principle
of the weak-phase solution is presented in  \cite{gonsalves2001}, 
but we found slight modifications \cite{keller2012spie}
leading to significantly better performance.
The algorithm uses focal-plane images and 
phase-diversity information to solve the wavefront, and the
estimated wavefront is corrected with a wavefront correcting
device. The correction step produces phase-diversity
information and a new image 
that are again used to compute the following phase
update. The schematic illustration of the algorithm is shown in 
Fig.~\ref{fg:algoscemaff}.

The obvious limitation of the FF algorithm is the assumption of the pupil
amplitudes being even. This holds reasonably well for most
of the optical systems having a circular shape, possibly with a
central obstruction. However, to achieve the optimal focal-plane
wavefront sensing with a high-order system not suffering from other
limiting factors, it is necessary to consider imaging models where the
pupil amplitudes can have an arbitrary shape. We have approached the problem by combining the FF-style weak-phase solution and a version of the Gerchberg-Saxton (GS) algorithm. The new algorithm is referred to as FF-GS in the following.

A detailed presentation and analysis of FF and FF-GS is shown in \cite{korkiakoski2014}.

\subsection{Phase retrieval with convex optimization}
\label{sec:conv}

A wave of new research \cite{candes2013phase, fogel2013phase, waldspurger2012phase, demanet2012stable} has emerged since it was shown in 2011 that a lifting technique called PhaseLift \cite{candes2013phase} can be used as a tool for phase retrieval. It uses tools from optimization theory to trade ("lift") a non-linear optimization problem into a well-known convex problem. However, the amount of unknown parameters increases dramatically.

The original idea of PhaseLift is to formulate the standard phase-retrieval problem as rank-minimization problem and then approximate it by minimizing the trace. However, if there exist a unique solution, the trace minimization is not necessary, and it is possible to solve the lifted problem as a convex feasibility search. Such a feasibility problem can be solved using POCS-style algorithms (projection onto convex sets), and the convergence is faster than PhaseLift, if the optimization is done using standard gradient search  \cite{demanet2012stable}.

The current literature, however, hardly says anything about results on data sets of any realistic size. To apply the lifting techniques for sufficiently large sets, it is necessary to introduce additional approximations to reduce the computation costs to feasible levels. Here, we show how a practical implementation of POCS-style codes can be  done for a lifted phase-retrieval case. The used notations are listed in Table~\ref{tb:sym}.

\begin{table}[hbtp] \begin{center}
\caption{Notations}
\label{tb:sym}
\begin{tabular}{ll}
  \hline
$\xph$ & unknown complex field (phase and amplitude),  a matrix of $N\times N$ \\
$\bm_i$ & $i$th measurement array, a matrix of $M \times M$ \\
$\xv$ & stacked vector of $\xph$ \\
$\bv_i$ & stacked vector of $\bm_i$ \\
$\aft_i$ & matrix modelling the linear relationship (FFT) from $\xv$ to $\bv_i$ \\
$\xlim$ & lifted unknown variables, a matrix of $N^2 \times N^2$ \\
$\xliv$ & stacked vector of $\xlim$ \\
$\phdm_i$ & $i$th phase diversity, a matrix of $N\times N$ \\
$\phdv_i$ & stacked vector of $\phdm_i$ \\
$\ali_i$ & matrix modelling the linear relationship from $\xliv$ to $\bv_i$ \\
  \hline
\end{tabular}\\
\end{center} \end{table}

The traditional phase retrieval problem can be written as
\begin{align} \label{eq:ph0}
  \bv_i = |\aft_i \xv|^2 = |\aft <\phdv_i, \xv>|^2,
\end{align}
where $\bv_i$ ($i=1,\ldots, M$) is the $i$th intensity measurement vector, $\xv$ is the unknown field and $\aft_i$ is the matrix that models the linear relationship between the phase and measurement domains. $\phdv_i$ is the electric field corresponding to the known phase diversity -- it is introduced between the measurements. In many imaging applications, the situation can be very accurately modeled by the Fraunhofer diffraction,  
\begin{align} \label{eq:ph0b}
  \bm_i = |\fft{<\phdm_i, \xph}>|^2,
\end{align}
where $\fft{\cdot}$ is a two dimensional Fourier transform and $<\cdot,\cdot>$ denotes an inner product. In this work, we focus on the Fraunhofer diffraction case, and therefore the matrices $\aft_i$ are defined through the use of point-wise multiplication of the phase diversities and the application of FFTs.

PhaseLift formulates the nonlinear inverse problem as a convex optimization problem:
\begin{align} \label{eq:ph}
  \min Tr(\xlim) & \\
  \alpha(\xlim) &= \bv \\
  \xlim &\succeq 0,
\end{align}
where $\xlim$ is a complex unknown, defined by lifting of $\xv$: $\xlim = \xv \xv^H$. $\alpha(\cdot)$ is a linear mapping from positive semidefinite matrices to a vector matching the measurement dimensions.

The PhaseLift problem is identical to the original phase retrieval problem, if a unique solution exist for $\xv$ \cite{demanet2012stable}. The trace minimization was introduced to approximate the rank minimization, which makes intuitively sense. However, given the unique solution, it is not necessary. PhaseLift can be solved using general convex optimization tools. However, if the trace minimization is omitted, the problem is reduces to a feasibility search, which can be solved faster.

If the PhaseLift is simplified into a feasibility problem, it can be solved using general algorithms like Projection Onto Convex Sets (or its faster extensions). The general idea is to project the current estimate (\xliv) alternately to the sets that satisfies 
\begin{equation} \label{eq:cs1}
  b_i = \ali_i \xliv   
\end{equation} 
and
\begin{equation} \label{eq:cs2}
  \xlim \succeq 0.
\end{equation}

To solve the problem with a practical dimension, some approximations are needed. For practical reasons, it is not possible to store the whole matrix of $\xlim$. Therefore, we represent it as
\begin{equation} 
\xlim = \sum_{j=1}^K \lambda_j x_j x_j^*,
\end{equation}
where $\lambda_j$ is the $j$th eigenvalue and $x_j$ is the $j$th eigenvector.

Thus, we maintain a factorized version of $\xlim$ at all times ($K$ largest eigenvalues are kept), and therefore the projection onto $\xlim \succeq 0$ becomes trivial: set the negative eigenvalues to zero. Also the recovery of the original unknown ($\xv$) is (in principle) easy: it is the eigenvector corresponding to the largest eigenvalue. The projection to satisfy the measurements could be done using the matrix equations as:
\begin{equation} 
  \xliv_{k+1} = \xliv_k - \ali_i^*(\ali_i^*\ali_i)^{-1}(\ali_i \xliv_k - \bv_i),
\end{equation}
where $\xliv_k$ is the $k$th estimate and $i$ is the measurement index, $i=1,\ldots K$. Due to the properties of the Fourier transform, this simplifies to 
\begin{equation} 
  \xliv_{k+1} = \xliv_k - \ali_i^*(\ali_i \xliv_k - \bv_i).
\end{equation}

However, the matrices $\ali_i$ are too large to be explicitly modeled, and we would like to use the FFTs to make the computation faster. To implement this efficiently, the latter part can be calculated as series of FFTs (as many FFTs as there are non-zero eigenvalues in $\xlim_k$),
\begin{equation} 
u_i =  \ali_i \xliv_k - \bv_i = \text{vec}\left( \sum_{j=1}^K \lambda_j |\ft{<\text{mat}(x_j), \phdm_i}>|^2\right) - b_i,
\end{equation}
where $\lambda_j$ is the $j$th eigenvalue of $\xlim_k$ and $\phdm_i$ is the field corresponding to the phase diversity of the $i$th image. The multiplication by $\ali_i^*$ is more difficult since it projects the vector of size $N$ (number of measurements) to the dimension of $N^2$. To find the approximation of this, we apply power methods on the quantity
\begin{equation} 
  Z = \xlim_{k+1} = \xlim_k - \text{mat}(\ali_i^*u_i).
\end{equation}
to find its $K$ highest eigenvalues. This requires finding a fast way to implement a matrix operation
\begin{equation} 
  Zv = \xlim_k v - \text{mat}(\ali_i^*u_i) v.
\end{equation}
This can be achieved by two consecutive FFTs and simple algebra,
\begin{align}
p_1 &= \fft{<\text{mat}(v), \phdm_i}>, \\
p_2 &= <\ifft{<p_1, \text{mat}(u_i)>}, 1/\phdm_i>,
\end{align}
and
\begin{align} \label{eq:xx}
  Zv &= \xlim_k v -   \text{vec}(p_2) \\
  &= \sum_{j=1}^K \lambda_j x_j (x_j^H v) - \text{vec}(p_2).
\end{align}
This linear operation ($Zv$) is then done sequentially as a part of power iteration to find the $K$ highest eigenvalues of $Z$, which is the next iterate for $\xlim$.

In practice, as pointed out by many others, the projection algorithms get often significantly faster, if some memory is introduced in the process. We define $Y=2\xlim_k - \xlim_{k-1}$, and projection is applied on $Y$ instead of $\xlim_k$. We also applied this modification.

\section{Results}
\label{sec:results}

Our earlier publications \cite{korkiakoski2012spie1, korkiakoski2013, korkiakoski2014} show in detail -- both simulations and laboratory experiments were carried out -- how the FF and FF-GS work. Here, we illustrate, with the help of numerical simulations, how the alternative approaches would perform. 

This section shows a comparison of the FF-GS algorithm with both the traditional GS algorithms and the convex optimization. As a test case, we use date from the setup we have used with our earlier publication \cite{korkiakoski2014}. It has optical aberrations typical for high-contrast adaptive optics; a spatial light modulator is used to correct those aberrations with the help of a static focal-plane camera.

\subsection{FF-GS compared with GS}

\begin{figure}[hbtp]  \center
\includegraphics[width=0.45\columnwidth]{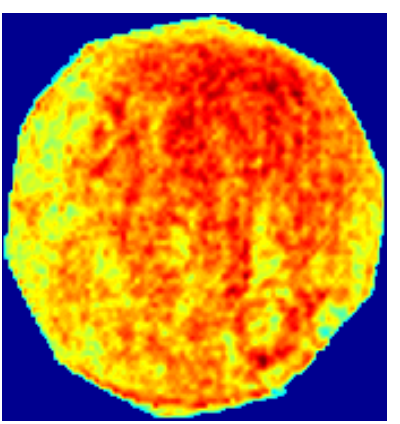}

\caption{Pupil amplitudes ($170\times 170$~pixels) used in the simulations.}
\label{fg:pupest1}
\end{figure}

We use numerical simulations that are shown to model the reality rather accurately \cite{korkiakoski2013,korkiakoski2014}. We model eight different sources of errors explicitly:
\begin{enumerate}
\item SLM quantification. We use only 6 bits to control the
  wavefront. 

\item PSF sampling. The wavefront and the resulting PSF are sampled
  internally by a factor of two higher than what the hardware controls
  or observes. The control algorithms use re-binned PSFs, and the
  simulated wavefront correction is interpolated bilinearly from
  the reconstruction at a resolution of $170\times 170$.

\item Image noise and dynamic range. We estimate the read-out noise of
  the HDR images to be at a level of $2.2\cdot 10^{-6}$ of the image
  maximum. Gaussian random noise is added to the simulated PSFs. The
  HDR images have maximum values $\sim$$4\cdot 10^8$,
  corresponding to about 29 bits, and this is also modeled in the simulations.

\item Background level. Standard background subtraction is performed
  on the PSF images, but a small error will still
  remain. Therefore, we add a constant background level,
  $2.7\cdot 10^{-6}$ of the image maximum, to the simulated PSFs.

\item Non-perfect pupil. Instead of the perfect top-hat function, we
  use pupil amplitudes shown in Fig.~\ref{fg:pupest1}. See also \cite{korkiakoski2014}.

\item Amplitude aberrations. We simulate the coupling of the wavefront
  and the transmission of the SLM as illustrated in \cite{korkiakoski2014}.

\item Alignment errors. Although the dOTF calibration is rather accurate,
  some error could still be present in the affine transform that we use to
  map the wavefront to the SLM pixels. The simulations indicate that
  if the transform has a mismatch corresponding to a rotation larger
  than 0.4$^\circ$, FF and FF-GS would be unstable. In practice, with
  the used hardware, we saw no hints these of
  instabilities. Therefore, a rotation error of 0.4$^\circ$
  represents the maximum misregistration that the wavefront control
  algorithms are likely to experience.

\item Tip-tilt error. Internal turbulence in the optical setup causes
  frame-to-frame wavefront variations, which can be approximated
  to a degree as small shifts of the recorded images.
  We measured the difference of the
  center-of-gravity between two consecutive PSFs recorded with the HDR
  method, and it was found to be on average 0.025~pixels.
  This error cannot be taken into account by the phase-diversity
  approach, and we model its impact on the performance.
  
\end{enumerate}

At first, we show how the conventional GS algorithm detects the iteratively corrected aberrations. We let FF algorithm to run 3--5 initial iterations, and the resulting new images and phase-diversities are fed to the GS algorithm to retrieve the phase. The retrieved WF is then iteratively used to correct the aberrations. To optimize the GS performance, we consider the following parameters:
\begin{enumerate}
\item Loop gain in the iterative correction. A value of 0.3 of optimal for FF-GS, and a value of 0.9 or higher is optimal for GS.

\item Number of used images. We test the algorithm with 3 and 5 images.

\item Number of GS iterations (alternate projections to all the images in the measurement set). A rough estimate of the WF is obtained with only 10 GS iterations. We also test a more accurate reconstruction with 50 iterations.

\item Restriction of the pupil amplitudes. We test the algorithm both in its traditional form (enforce even pupil amplitudes over the aperture) and in an unrestricted more where no pupil-plane amplitude restrictions are enforced.
\end{enumerate}
 
\begin{figure}[hbtp]  \center
\includegraphics[width=0.49\columnwidth]{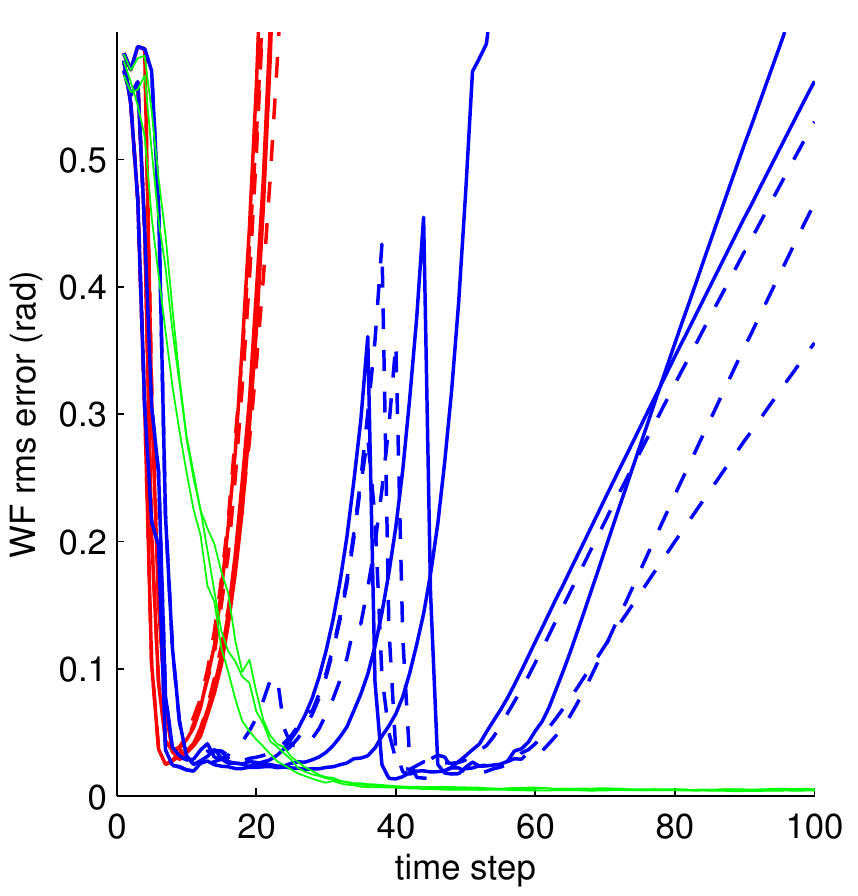} 
\includegraphics[width=0.49\columnwidth]{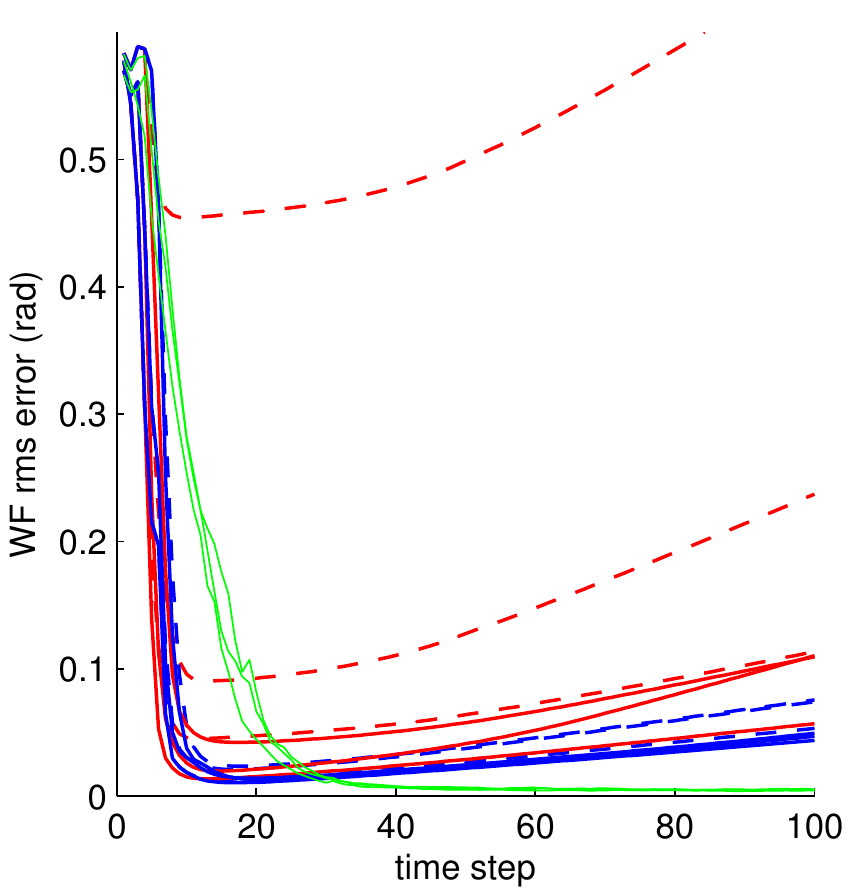} 

\includegraphics[width=0.49\columnwidth]{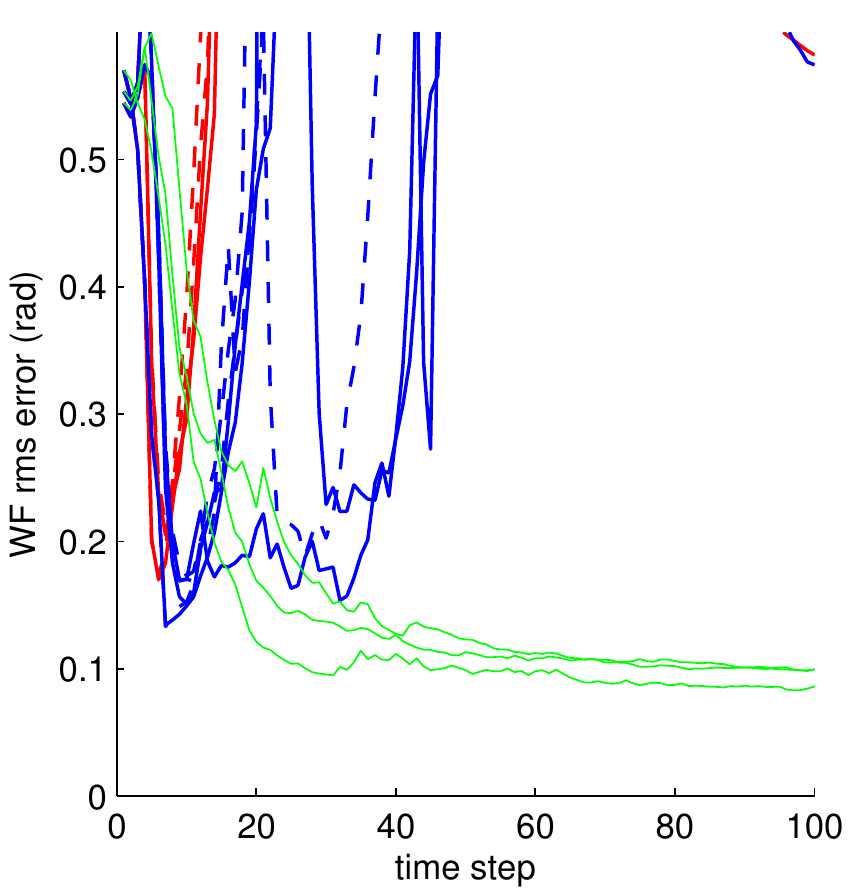}
\includegraphics[width=0.49\columnwidth]{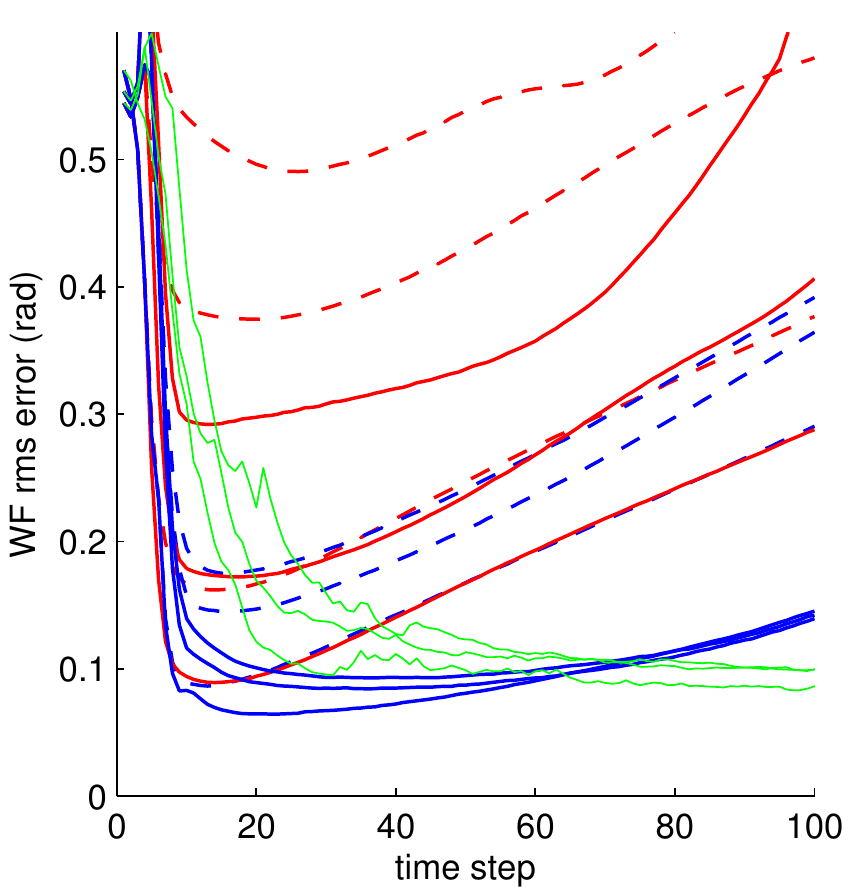}
\caption{Comparison of GS and FF-GS algorithms: WF rms error as a function of time step. Upper row: error free case. Lower row: all errors simulated. Left column: GS restrict pupil amplitudes. Right column: GS lets pupil amplitudes have arbitrary values. Green lines: FF-GS for comparison, data from \cite{korkiakoski2014}. Blue lines: GS uses 5 images. Red lines: GS uses 3 images. Dotted lines: 10 GS iterations. Solid lines: 50 GS iterations.}
\label{fg:gscmp}
\end{figure}

Fig.~\ref{fg:gscmp} shows the results. The upper row has the case where simulations are down with no error sources. The lower row has the case with all the error sources modeled. 

It appears to be very difficult to make the GS algorithm stable. The best results are obtained with 5 images and 50 GS iterations (possibly further images and projection iterations would still improve the results). A lower amount used images or GS iterations works to some extent without error sources if the pupil amplitudes are restricted. If pupil amplitudes are not restricted (thus also an estimate for the pupil amplitudes is obtained), five images are needed also in the error-free case -- but no significant gain is obtained by using 50 instead of 10 GS iterations. The more practical simulations show successful results only with the higher number of GS iterations and five used images. 

In total about ten wavefront correcting iterations are needed to reach close to the optimum. Without error sources, $\sim$15 iterations produce the best results, with the error sources 20--30 iterations are needed. At its optimum, the GS algorithm with five used diversity images has $\sim$20\% higher residual WF error compared to FF-GS, but in the error-included simulations the WF error is $\sim$20\% lower. However, the comparison to FF-GS is unfair in the sense that it uses only three images as implemented in \cite{korkiakoski2014}.

After the optimum, the GS algorithm becomes unstable, and the wavefront error starts to creep upwards. This happens since the phase-diversity becomes too small to give useful extra information. In addition, unlike in the FF-GS, the phase-diversity information is not enforced when applying the image-plane restrictions.

\begin{figure}[hbtp]  \center
\includegraphics[width=0.32\textwidth]{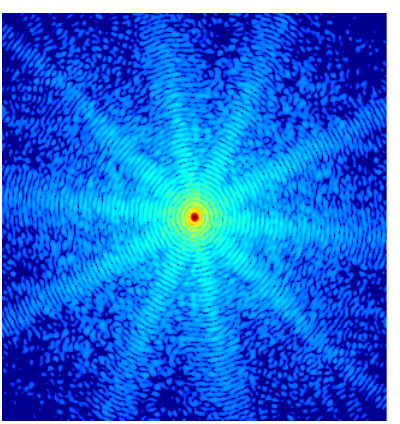}
\includegraphics[width=0.32\textwidth]{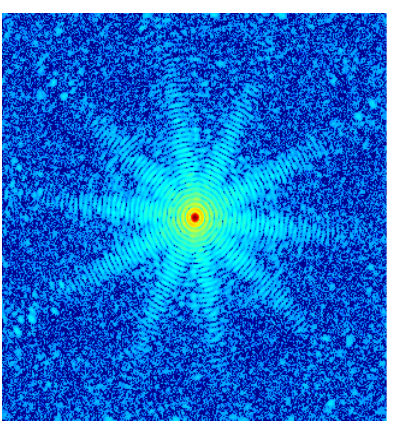}
\includegraphics[width=0.32\textwidth]{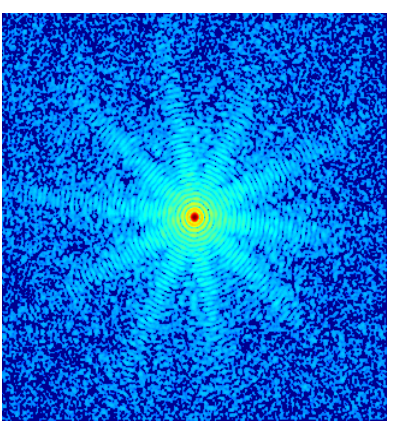}

\includegraphics[width=0.32\textwidth]{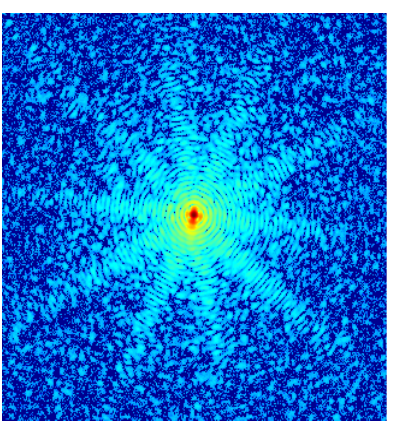}
\includegraphics[width=0.32\textwidth]{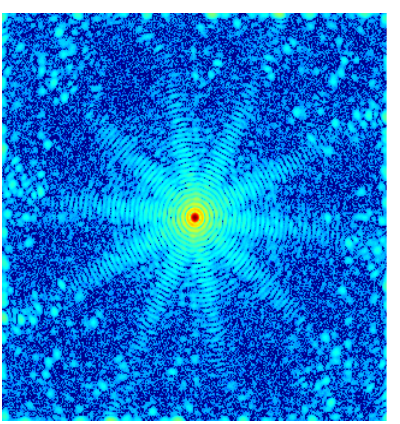}
\includegraphics[width=0.32\textwidth]{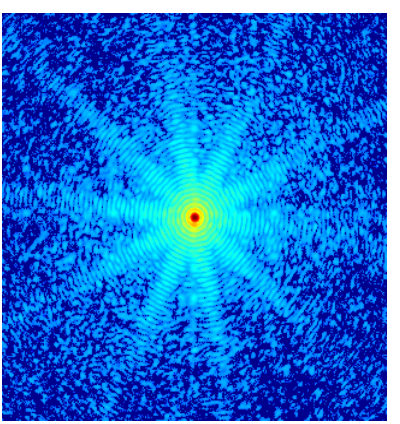}
\caption{Examples of the focal-plane images raised to 0.1 power. Top row: perfect image (no wavefront aberrations), GS algorithm with optimal parameters after 25 iterations,   FF-GS algorithm converged (simulated).  Bottom row: initial image, GS algorithm with optimal parameters after 100 iterations, FF-GS algorithm converged (laboratory measurement \cite{korkiakoski2014}). }
\label{fg:p1gs}
\end{figure}

The creep appears as an amplification of certain higher spatial frequencies. Those frequencies are visible as single speckles, and each iterations makes them brighter. We illustrate this in Fig.~\ref{fg:p1gs}. The top-left image shows the ideal image without any wavefront aberrations (the visible aberrations are caused by non-uniform transmission). The optimal GS algorithm (5 images, 50 GS iterations, 25 correction iterations) has corrected well all the lower spatial frequencies, but a few uncorrected speckles remain further on the halo. After total 100 correction iterations those uncorrected speckles have become brighter and many more have emerged. 

A visual inspection also reveals that most of the problematic speckles appear also in the initial uncorrected image. Thus, the GS algorithm fails to be reliable at the high spatial frequencies. The last column shows results of the FF-GS algorithm for comparison. FF-GS has no problematic speckles that would grow when repeating the correction iterations, which makes it significantly more robust compared of the GS approach. The overall WF error, however, is $\sim$20\% higher compared to the ideal GS performance -- but this could likely be improved if the algorithm was updated to use as many images as the GS method is using. FF-GS used three images while the best GS realization here used five.

\subsection{Results with convex optimization}

Here, we test the new idea of convex optimization (see Section~\ref{sec:conv}) applied in phase retrieval. The test case is the same as in the previous section, but we show the results only for the error-free case. We use five images and corresponding diversity information.

However, even with the fast implementation we wrote, the implementation is still too low to effectively use it as an iterative corrector. Therefore, we do only a single phase retrieval and compare the results to the expected values. The comparison to the previous sections is therefore not entirely fair since there we used several correction iterations to improve the performance.

\begin{figure}[hbtp]  \center
\includegraphics[width=0.32\textwidth]{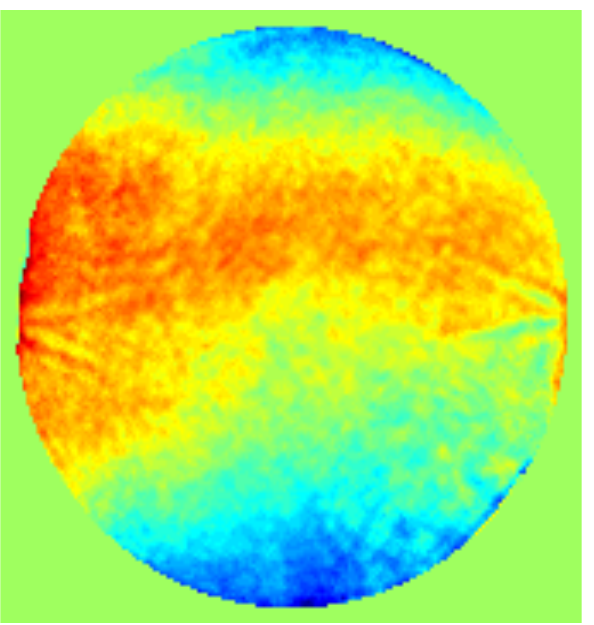}
\includegraphics[width=0.32\textwidth]{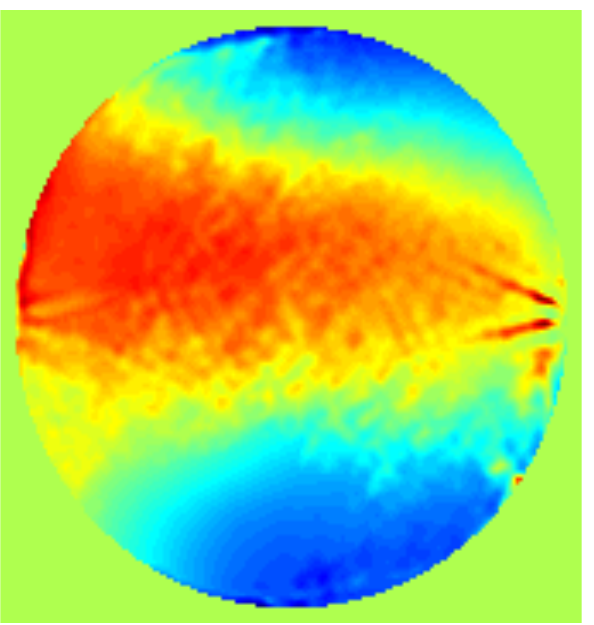}
\includegraphics[width=0.32\textwidth]{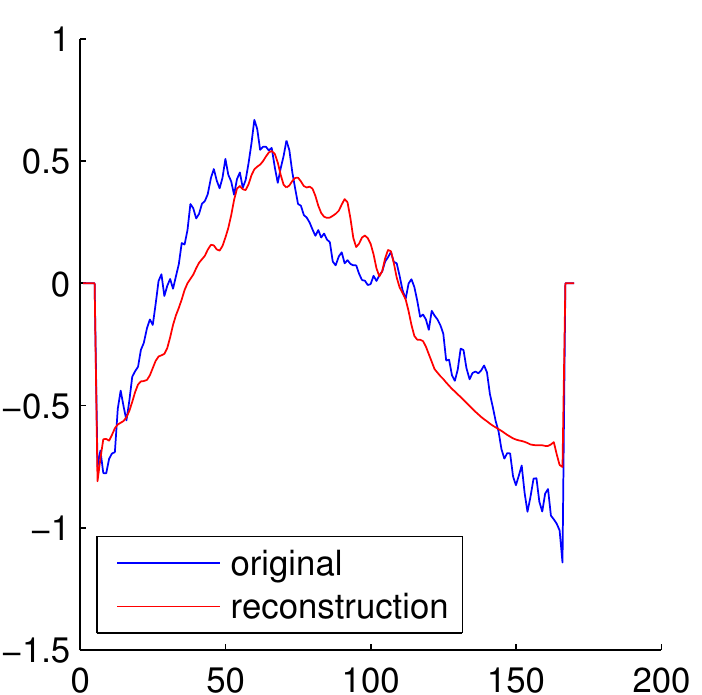}
\caption{Results for phase retrieval with convex optimization. Left: the unknown WF to be reconstructed.  Middle: the reconstruction. Right: radial cuts of the images (units are in radians).}
\label{fg:wfs}
\end{figure}

\begin{figure}[hbtp]  \center
\includegraphics[width=0.5\textwidth]{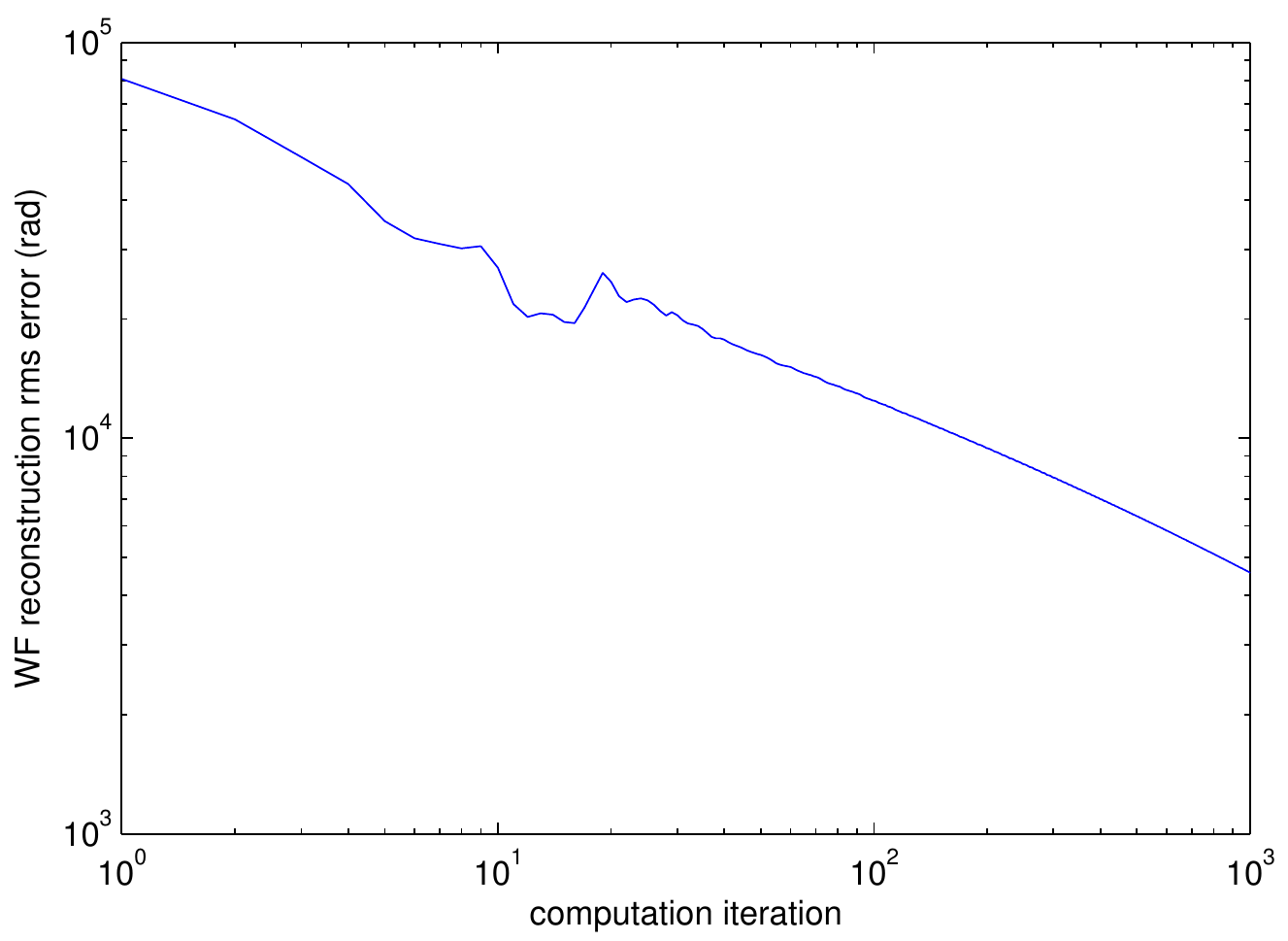}
\caption{Convergence of the POCS algorithm for phase retrieval.}
\label{fg:conv}
\end{figure}

Fig.~\ref{fg:wfs} shows the original wavefront and the final reconstruction of the POCS algorithm. The figure also shows radial cuts of the wavefronts. The convergence properties are shown in Fig.~\ref{fg:conv}. We see that 1000 iterations are not enough for a sufficient convergence since the convergence becomes extremely slow after 10--100 first iterations. Each iteration requires about 200 FFTs with a dimension of $640\times 640$. Total calculation time for the final reconstruction took several hours with a modern PC and standard Matlab FFT implementation.

We note that the POCS implementation is well able to retrieve the low spatial frequencies of the wavefront, but it fails to correctly find the sharp high-order features. In theory, this might still happen if the algorithm is let to converge -- but this option is still limited by the computational requirements. It is also possible that iterative correction, in the same way as with the FF-GS and GS algorithms, would help to correctly retrieve the unknown wavefront. However, the computational problems make also this approach difficult.

\begin{figure}[hbtp]  \center
\includegraphics[width=0.49\textwidth]{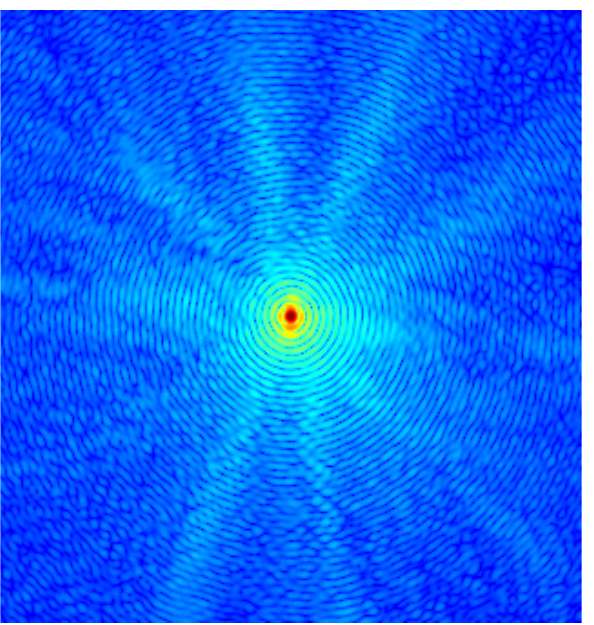}
\includegraphics[width=0.49\textwidth]{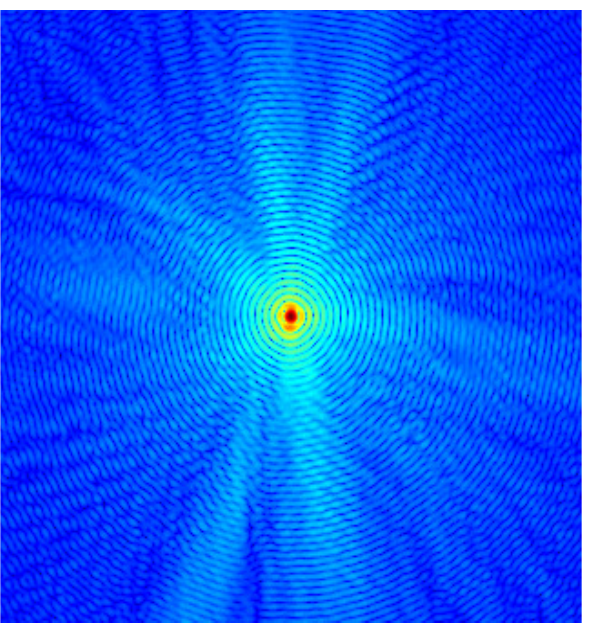}
\caption{Illustration the performance of our POCS based phase retrieval. Left: recorded image (no diversity). Right: the image resulting from retrieved wavefront.}
\label{fg:cofigs}
\end{figure}

We also show here, in Fig.~\ref{fg:cofigs}, the PSF image of the unknown wavefront and the PSF resulting from the estimated wavefront. The image also suggests that the high spatial frequencies are not correctly estimated, and the dataset should have enough information to yield a better wavefront estimate.

\section{Conclusions and discussion}
\label{sec:conclusions}

We have studied different options to use focal-plane wavefront sensing to determine the unknown pupil-plane field at a very high spatial resolution (more than 150$\times$150 pixels). With the help of numerical simulations, we tested the performance of traditional Gerchberg-Saxton, recently proposed Fast \& Furious algorithm and a convex optimisation based phase retrieval.

Our results indicate that the methods based on fast iterative correction and linear approximations tend be easier to implement in a reliable way. The FF and FF-GS algorithms have been shown to an excellent choice for this type of phase-retrieval \cite{korkiakoski2014}.

The GS method works to some extent, but, in particular at high spatial frequencies, it is difficult to find reliable wavefront estimates. In addition, its computational requirements are $\sim$100 times higher compared to the FF-GS.

The phase retrieval based on convex optimisation also works to some extent, but our numerical implementation is not fast enough to make it a competitive alternative to the traditional methods.


\bibliography{../report}   
\bibliographystyle{spiebib}   

\end{document}